\documentclass[aps,prl,twocolumn,superscriptaddress,showpacs,preprintnumbers]{revtex4}

\usepackage{graphicx}

\begin{document}

\title{Quasiparticle-like peaks, kinks, and electron-phonon coupling
at the ($\pi$,0) regions in the CMR oxide
La$_{2-2x}$Sr$_{1+2x}$Mn$_{2}$O$_{7}$}

\author{Z. Sun}
\affiliation{Department of Physics, University of Colorado, Boulder,
CO 80309, USA}
\affiliation{Advanced Light Source, Lawrence Berkeley
National Laboratory, Berkeley, CA 94720, USA}
\author{Y. -D. Chuang}
\affiliation{Advanced Light Source, Lawrence Berkeley National
Laboratory, Berkeley, CA 94720, USA}
\author{ A. V. Fedorov}
\affiliation{Advanced Light Source, Lawrence Berkeley National
Laboratory, Berkeley, CA 94720, USA}
\author{J. F. Douglas}
\affiliation{Department of Physics, University of Colorado, Boulder,
CO 80309, USA}
\author{D. Reznik}
\affiliation{Forschungszentrum Karlsruhe, Institut f\"{u}r
Festk\"{o}rphysik, Postfach 3640, D-76021 Karlsruhe, German}
\author{F. Weber}
\affiliation{Forschungszentrum Karlsruhe, Institut f\"{u}r
Festk\"{o}rphysik, Postfach 3640, D-76021 Karlsruhe, German}
\author{N. Aliouane}
\affiliation{Hahn-Meitner-Institut, Glienicket Str 100, Berlin
D-14109, Germany}
\author{D. N. Argyriou}
\affiliation{Hahn-Meitner-Institut, Glienicket Str 100, Berlin
D-14109, Germany}
\author{H. Zheng}
\affiliation{Materials Science Division, Argonne National
Laboratory, Argonne, IL 60439, USA}
\author{J. F. Mitchell}
\affiliation{Materials Science Division, Argonne National
Laboratory, Argonne, IL 60439, USA}
\author{T. Kimura}
\altaffiliation{Present address: Los Alamos National Laboratory, Los
Alamos, NM 87545, USA}
\affiliation{Department of Applied Physics,
University of Tokyo, Tokyo 113-8656, Japan}
\author{Y. Tokura}
\affiliation{Department of Applied Physics, University of Tokyo,
Tokyo 113-8656, Japan}
\author{A. Revcolevschi}
\affiliation{Laboratoire de Physico-Chimie de l'Etat Solide,
University of Paris Sub-11, 91405 Orsay, Cedex, France}
\author{D. S. Dessau}
\altaffiliation{To whom correspondence should be addressed:
Daniel.Dessau@colorado.edu}
\affiliation{Department of Physics,
University of Colorado, Boulder, CO 80309, USA}

\date{\today}

\begin{abstract}

Using Angle-Resolved Photoemission (ARPES), we present the first
observation of sharp quasiparticle-like peaks in a CMR manganite. We
focus on the ($\pi$,0) regions of k-space and study their electronic
scattering rates and dispersion kinks, uncovering the critical
energy scales, momentum scales, and strengths of the interactions
that renormalize the electrons.  To identify these bosons we
measured phonon dispersions in the energy range of the kink by
inelastic neutron scattering (INS), finding a good match in both
energy and momentum to the oxygen bond-stretching phonons.
\end{abstract}


\maketitle

The exotic physics in condensed matter systems, such as high
temperature superconductivity in cuprates \cite{Bednorz} and
colossal magnetoresistance (CMR) in manganites
\cite{Tokura,Chatterji}, is due to strong many-body interactions of
unknown origin. Relevant interactions are either electron-electron
or electron-boson, where the boson is a collective excitation such
as a magnon or, as in the case of conventional superconductivity
\cite{Schrieffer}, a phonon.  These interactions are typically
described as "dressing" the electrons to create quasiparticles, the
properties of which determine physical quantities such as electrical
and thermal conductivity. In certain cases such as one-dimensional
"Luttinger Liquids" \cite{Voit} and possibly cuprates
\cite{Anderson} and manganites \cite{Dessau}, the electronic
correlations are so strong that the quasiparticle concept falls
apart, and only broad and strongly damped electronic excitations
have been observed. A side effect of this is that the absence of the
quasiparticles gives experimentalists fewer windows into the
interactions responsible for the behaviour of the system.

Many-body interactions in manganites are expected to be strong and
include the coupling to other electrons plus collective modes such
as phonons \cite{Perebeinos,Millis}, magnons \cite{Jaime,Hirota},
and orbitons \cite{van,Saitoh} . Although some of these bosonic
modes have been extensively studied in the manganites through Raman,
X-ray and neutron scattering experiments, the details of how these
modes couple to the electrons have been almost completely
unexplored.  In essence we do not know which modes are most relevant
or how strongly they couple to the electrons.

La$_{2-2x}$Sr$_{1+2x}$Mn$_{2}$O$_{7}$ is a naturally layered
compound with two MnO$_{2}$ planes per unit cell, with the physical
properties dominated by these bilayers \cite{Mitchell}. The
x=0.36-0.4 samples we studied have a transition from a high
temperature paramagnetic insulating (PI) state to a low temperature
ferromagnetic metallic (FM) state at Tc$\sim$120-130K. To uncover
the properties of electrons and their interactions with bosonic
modes in these compounds, we take advantage of two powerful energy
and momentum-resolved techniques, ARPES and inelastic neutron
scattering (INS) to individually probe the electrons and bosons
throughout the zone, respectively. This powerful combination gives
unprecedented clarity into the many-body interactions in the CMR
compounds. We observed sharp quasiparticle-like peaks using ARPES,
which opens a window for us to study electronic dispersions and
scattering rates and how electrons are renormalized by bosonic
modes. Phonon dispersions were measured by INS, which have a good
match in both energy and momentum to that which should couple to
electrons, indicating the relevance of phonons to the interactions
in manganites.

The ARPES experiments were performed at beamline 12.0.1 of the
Advanced Light Source, Berkeley, using a Scienta SES100 electron
spectrometer under a vacuum better than 3$\times$10$^{-11}$ torr.
All samples were cleaved \emph{in-situ} at 20 K, and low energy
electron diffraction (LEED) patterns confirmed the high quality of
the surfaces. The combined instrumental energy resolution of
experiments was better than 20 meV and the momentum resolution was
about 2 \% of the zone edge (.02 $\pi$/a). The neutron scattering
experiments were performed on the 1T triple axis spectrometer at the
ORPHEE reactor at Saclay utilizing the 220 reflection of copper as
the monochromator and the 002 reflection of Pyrolytic graphite. The
sample was mounted in a closed-cycle refrigerator with the
measurements performed at 11K.

The electronic spectral function is determined from ARPES data. Fig
1a illustrates the typical Fermi surface of
La$_{2-2x}$Sr$_{1+2x}$Mn$_{2}$O$_{7}$ \cite{Dessau}. Theoretically
there is a small piece at the zone center consisting of primarily
out-of-plane d$_{3z^2-r^2}$ Mn-O states, while the large hole
pockets centred at the zone corners are due to bilayer-split
in-plane d$_{x^2-y^2}$ Mn-O states \cite{Dessau}. These in-plane
states, which are expected to be more important for the transport
and magnetic properties of these layered materials, are the focus of
this paper. All previous work had been unable to resolve the bilayer
splitting. In this work we discovered we could selectively pick up
either the bonding or antibonding bands near ($\pi$,0) regions by
tuning the matrix elements, which allows a highly accurate analysis
of data. We empirically found that bonding and antibonding portions
of the bilayer-split bands are emphasized with 73 eV and 56 eV
photons, respectively. The spectral weight at the Fermi energy taken
using 56 eV photons at T=20K is shown in fig 1b, giving a
(matrix-element modulated) experimental mapping of the 2D Fermi
surface (FS). The weak feature around the zone center as indicated
by the red dots in fig 1b is consistent with the prediction of the
small electron pocket. Here we focus on data along the blue line in
panel a, i.e. the line (k$_x$, 0.9$\pi$/a). Fig 1c shows a wide
binding energy scan of the antibonding band as a function of k$_x$
at T=20K, exhibiting a clear parabolic dispersive feature with
maximal spectral weight around the binding energy 0.4 eV. The
bonding band (not shown) has a similar dispersive feature and
reaches 400 meV deeper binding energy. The black dots superimposed
on the right half of the data show this wide-scale dispersion
clearly. Near E$_F$ (below the red arrow) the dispersive feature is
sharp and heavily renormalized from the parabolic dispersion,
indicating important many-body effects.

An Energy Distribution Curve (EDC) taken at k=k$_F$ from this data
(vertical white line in panel c) is plotted in fig 1d. Near E$_F$
there is a clear and well-resolved quasiparticle-like peak \cite{qp}
- the first such observation in a CMR oxide. This observation by
itself has important ramifications for the study of electronic
correlations in low dimensional systems such as the layered
manganites, as certain important classes of models of correlated
electrons \cite{Voit,Anderson} require the absence of such
quasiparticles. Compared to data on
La$_{2-2x}$Sr$_{1+2x}$Mn$_{2}$O$_{7}$ (x=0.4) which have very small
spectral weight near E$_F$ \cite{Dessau,Chuang,peak}, the appearance
of a quasiparticle-like peak is not likely due to improved sample
quality issues \cite{peak} but rather due to the reduced level of
AF-canting \cite{Kubota} in the x= 0.36 and 0.38 samples, which
tends to make these samples electronically more three-dimensional.
This trend is consistent with that observed in the cobaltates
\cite{Valla} and ruthenates \cite{Wang} , in which it has been
argued that a high dimensionality favors more quasiparticle spectral
weight. The overall EDC lineshape shows a peak-dip-hump structure
(panel d). In the theoretical framework of a Fermi liquid, the hump
is the incoherent part of the single particle excitation and can be
considered to be "shakeups" of bosonic excitations
\cite{Dessau,Perebeinos}.

\begin{figure}[tbp]
\begin{center}
\includegraphics[width=1\columnwidth,angle=0]{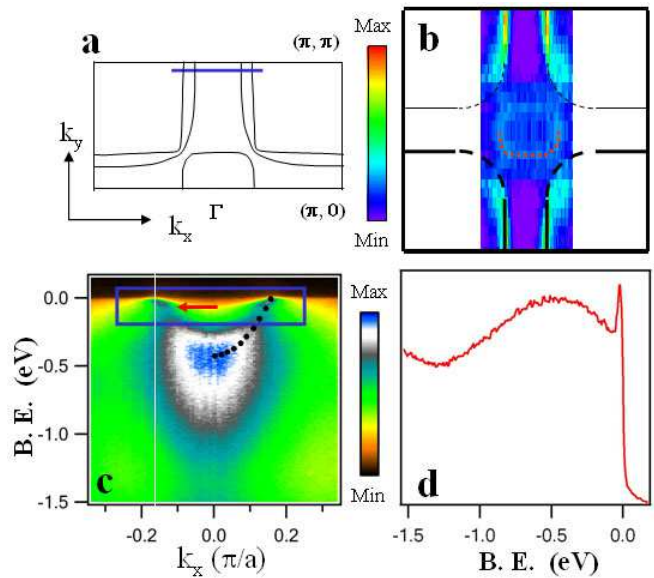}
\end{center}
\caption{Low-temperature (T=20 K) ARPES data from
La$_{2-2x}$Sr$_{1+2x}$Mn$_{2}$O$_{7}$ (x=0.38). (a) A representative
Fermi surface, after ref \cite{Dessau}. (b)The spectral weight at
E$_F$ over much of the first Brillouin zone and Fermi surfaces of
antibonding band (black lines) and d$_{3z^2-r^2}$ Mn-O states (red
dots). (c) Binding energy versus momentum ($\pi$/a) image plot of
the antibonding band from the 0.9 $\pi$ slice (blue line in panel
a), with dots determined from a combination of fits to EDCs (deeper
energies) and MDCs (near E$_F$). (c)The EDC at k$_F$ indicated by
the vertical white line in (c).}
\end{figure}

The clear presence of quasiparticle-like peaks in the current sample
gives us, for the first time, a new and detailed window into the
electronic correlations in the manganites. Figure 2 shows details of
the near-Fermi energy region of the data.  Panel a shows the
antibonding data as a function of k$_x$.  The blue points of Panel b
show the dispersion relation, which were determined from fits of
Momentum Distribution Curves (MDCs) with a Lorentzian lineshape on
top of a small monotonically varying background, while the red
points show the results from a similar analysis of bonding band data
\cite{sample} (upper axis - note the different scale from the bottom
axis). An s-shaped kink structure, deviating from the
non-interacting parabolic dispersion (black dots), can be clearly
seen for both bands. These deviations are due to the many-body
effects, and in the language of many-body physics are due to the
real part of the electron self-energy, Re$\Sigma$ . This kink
structure has been observed in x=0.4 samples before
\cite{Chuangchapter}, though we have better statistics now. The
slopes of the renormalized and non-interacting dispersions near
E$_F$ give the renormalized and bare Fermi velocities, respectively.
In a simple electron-boson coupling model, their ratio can be
parameterized as 1+$\lambda$, where $\lambda$ is a measure of the
electron-boson coupling strength (also termed the mass enhancement).
Since the ratio of the low energy slopes is approximately 2, this
would imply $\lambda$$\sim$1 for both bonding and antibonding bands.
We also note that specific-heat data has indicated similar values of
the coupling strengths \cite{Okuda,Woodfield} , though this analysis
is much less direct as it is based on a comparison to theoretical
band structure data. Hence this is the first direct information of
coupling parameters in manganites.

\begin{figure}[tbp]
\begin{center}
\includegraphics[width=1\columnwidth,angle=0]{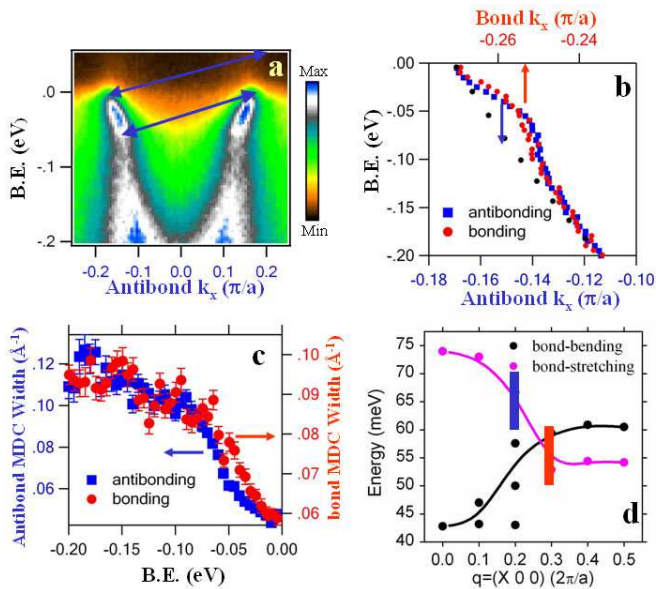}
\end{center}
\caption{Electron and phonon dispersion relations and scattering
rates. (a) Energy versus momentum image plot of the antibonding
electron band from the 0.9 $\pi$ slice (blue frame in figure 1c).
(b) MDC derived E vs. k dispersion of antibonding states (blue) and
bonding states (red) compared to a parabolic fit to the deeper lying
antibonding dispersion (black dots). (c) MDC full widths versus
energy for antibonding (blue) and bonding states (red). (d) Phonon
dispersion relations from neutron scattering from the bond
stretching (pink) and bond bending modes (black) of even symmetry.
The bond-bending vibration contributes to multiple modes at some
wavevectors because it mixes with other vibrations. The lines are
guides to the eye.  Electron kink scales and nesting q vectors for
the antibonding (blue) and bonding band (red) are also included. }
\end{figure}

Typically, the low temperature metallic state of the manganites is
considered to be a rather standard metal without strong correlation
effects.  The coupling strength  $\lambda$ of the order of unity
discovered here tells a different story however.  This is in the
intermediate-to-strong coupling regime \cite{Mahan}, and is on the
precipice such that a minor perturbation in parameters giving a
small increase in the coupling strength may lead to polaronic
localization. Therefore these couplings may likely be responsible
for the metal-insulator transition which is at the heart of the CMR
problem.

The identity of the important boson mode(s) can be determined from
these dispersions. Based on the maximum in Re$\Sigma$  at about
50-60 meV, we estimate the critical energy of the boson mode(s) is
around 60-70 meV for the antibonding band and about 50-60 meV for
the bonding band \cite{energyscale}. A clear step-increase in the
scattering rates centred at the same energy scales are observed
(fig. 2c). This self-consistency gives great confidence in the
assignment of the critical energy scales to the data. Modes at other
energy scales will of course have some relevance as well, but as
evidenced by the kink data they will couple to the electrons less
strongly than modes near 60 meV.

Various proposals for the important mode coupling in the manganites
have been made, including phonons \cite{Perebeinos,Millis}, magnons
\cite{Jaime,Hirota}, and orbitons \cite{van,Saitoh}.  A key point is
that the strongly nested Fermi surface should be highly susceptible
to a mode with a momentum transfer equal to the nesting vector
q$\sim$0.17$\times$(2$\pi$/a,0) for the antibonding band and
q$\sim$0.27$\times$(2$\pi$/a,0) for the bonding band. The blue arrow
in fig. 2a indicates the corresponding electron scattering within
the antibonding band. At these q vectors the excitation energy of
magnons is $\sim$ 30 meV \cite{Hirota}, and orbitons are larger than
100 meV over the entire zone\cite {Saitoh} and so are in
disagreement with the critical energy scales we observed here.
However, there are longitudinal optical phonons that couple to
charge fluctuations in the Mn-O layers exactly in this momentum and
energy range, which makes these phonons a good candidate for the
dominant coupling.

We performed neutron scattering measurements on
La$_{2-2x}$Sr$_{1+2x}$Mn$_{2}$O$_{7}$ (x=0.4) to determine the
phonon structures and dispersion relations of the longitudinal
bond-stretching and bond-bending phonons, as shown in figure 2d.
Both even and odd modes with respect to the bilayer structure of
La$_{2-2x}$Sr$_{1+2x}$Mn$_{2}$O$_{7}$ have been measured. Only even
phonons are shown here - the odd phonons have similar dispersions
but are a few meV lower. Overlaid with the phonon dispersion curves
on this plot are the kink energy scales from the ARPES data, plotted
at the respective q vectors where energy-momentum conservation
allows the phonons to scatter from intraband transition
(A$\rightarrow$A (blue), and B$\rightarrow$B (red)).  Interband
transitions such as A$\rightarrow$B will show up at an intermediate
position and are not included on the graph.  It is seen that both
the kink energies and nesting vectors of the electrons match closely
with the phonon energies and q values, giving high confidence that
it is these phonons which have dominant coupling to the electrons.
Especially important are the bond-stretching phonons, which should
couple most strongly to the bonding and antibonding electrons (the
bond-bending phonons may couple to the bonding electrons as well).
The downward dispersion of the bond-stretching phonons also is seen
to match well with the electron kink data, and is in contrast with
that found for the upward-dispersive magnons \cite{Hirota}.  In
fact, the downward dispersion of the bond-stretching phonons has
long been considered an anomaly in manganites and other perovskites,
as a simple shell-model predicts that the bond-stretch phonons
should have an upward dispersion \cite{Reichardt}.  The downward
dispersion is seen to occur at the same q values where the electron
nesting occurs, so future studies might consider whether the
coupling to the electrons renormalizes these phonon properties as
well.

The coupling of electrons to the bond-stretching phonon branch has
also been considered to be important in the cuprates. It has been
argued that this branch, which has a similar E vs. q relation, may
be responsible for electronic dispersion kinks near 70 meV
\cite{Lanzara}, especially in the nodal or ($\pi,\pi$) direction
(the antinodal states in the cuprates couple more strongly to a
lower energy scale mode such as the B$_1g$ phonon \cite{Cuk} or the
magnetic resonance mode \cite{Gromko}, though the latter coupling is
only apparent in the superconducting state). In the cuprates the q
value that connects the nodal states does not closely match the q
values of the bond-stretch phonon as determined from neutron
scattering. This is perhaps the reason why the coupling is in
general weaker in the cuprates than in the manganites. The studies
of electron-phonon coupling in manganites may provide opportunities
to uncover the roles phonons play not only in the CMR effect but
also in the pairing of electrons in high-T$_c$ cuprate
superconductors.

In summary, we observed sharp quasiparticle-like peaks in
La$_{2-2x}$Sr$_{1+2x}$Mn$_{2}$O$_{7}$ using ARPES, which allows us
to study how electrons are renormalized by boson modes in manganite.
Phonon dispersions were measured by INS, and the bond-stretch
phonons have a dispersion closely matching the electronic kinks in
both energy and momentum, indicating the important coupling of this
phonon branch to electrons in manganites.

We are grateful to T. Devereaux, T. Egami, N. Furukawa, and J. Zhang for helpful
discussions. This work was supported by the U.S. Department of Energy under grant
DE-FG02-03ER46066 and Contract W-31-109-ENG-38, and by the U.S. National Science
Foundation grant DMR 0402814.  The ALS is operated by the Department of Energy,
Office of Basic Energy Sciences.


\begin{thebibliography}{99}
\bibitem{qp}We call this peak quasiparticle-like because it is not yet
clear if all of the conditions set forth by Landau for true quasiparticles are met,
e.g. the peak width scaling quadratically with energy and temperature.
\bibitem{peak}We have also measured many La$_{2-2x}$Sr$_{1+2x}$Mn$_{2}$O$_{7}$ (x=0.4)
 samples on our same experimental setup, and in contrast to the x=0.38 samples we do
 not observe any signs of quasiparticles at any temperatures, photon energies, or
 polarizations.  Therefore the difference between x=0.38 and 0.4 appears to be due
 to intrinsic issues.
\bibitem{sample} Bonding band data was from an 0.36 doped sample from which we
obtained better statistics, though the same energy scale as from 0.38 data.
\bibitem{energyscale} In the limit of zero broadening we expect the kink energy to
asymptotically approach the boson energy scale from below.  In the
real case with finite signal strength and peak broadening, the kink
will never fully approach the boson energy scale and so the kink
energy will slightly underestimate the mode energy. We therefore
consider the kink energy $\sim$ 60 meV to be the lower limit of the
boson energy scale for the antibonding band  and $\sim$ 50 meV to be
the lower limit for the antibonding band.  The upper limit of the
boson scale should be about 10 meV greater, where we can clearly
distinguish a lone higher energy feature.
\end{thebibliography}
\end{document}